\documentclass[aip,pop, numerical, reprint,floats]{revtex4-1}
\usepackage{amssymb}
\usepackage{amsmath}
\usepackage{graphicx}
\usepackage{ulem}
\usepackage[latin1]{inputenc}
\usepackage{url}
\usepackage{graphicx}
\usepackage{color}

\begin{document}

\title{Kinetic modeling of divertor heat load fluxes in
the Alcator C-Mod and DIII-D tokamaks
}
\author{A.Y. Pankin}
\affiliation{Tech-X Corporation, Boulder, CO, USA}
\author{T. Rafiq}
\affiliation{Department of Physics, Lehigh University, Bethlehem, PA, USA}
\author{A.H. Kritz}
\affiliation{Department of Physics, Lehigh University, Bethlehem, PA, USA}
\author{G.Y. Park}
\affiliation{WCI Center for Fusion Theory, NFRI, Korea}
\author{C.S. Chang}
\affiliation{Princeton Plasma Physics Laboratory, Princeton, NJ, USA}
\author{D. Brunner}
\affiliation{MIT Plasma Science and Fusion Center, Cambridge, MA, USA}
\author{R.J. Groebner}
\affiliation{General Atomics, San Diego, CA, USA}
\author{J.W. Hughes}
\affiliation{MIT Plasma Science and Fusion Center, Cambridge, MA, USA}
\author{B. LaBombard}
\affiliation{MIT Plasma Science and Fusion Center, Cambridge, MA, USA}
\author{J.L. Terry}
\affiliation{MIT Plasma Science and Fusion Center, Cambridge, MA, USA}
\author{S. Ku}
\affiliation{Princeton Plasma Physics Laboratory, Princeton, NJ, USA}
\date{\today}
\begin{abstract}
The guiding-center kinetic neoclassical transport code, XGC0, [C.S. Chang \textit{et. al}, Phys.~Plasmas \textbf{11}, 2649 (2004)] is used to compute the heat fluxes and the heat-load width in the outer divertor plates of Alcator C-Mod and DIII-D tokamaks. The dependence of the width of heat-load fluxes on neoclassical effects, neutral collisions and anomalous transport is investigated using the XGC0 code.  The XGC0 code includes realistic X-point geometry, a neutral source model, the effects of collisions, and a diffusion model for anomalous transport. It is observed that width of the XGC0 neoclassical heat-load is approximately inversely proportional to the total plasma current $I_{\rm p}$. The scaling of the width of the divertor heat-load with plasma current is examined for an Alcator C-Mod discharge and four DIII-D discharges.  The scaling of the divertor heat-load width with plasma current is found to be weaker in the Alcator C-Mod discharge compared to scaling found in the DIII-D discharges. The effect of neutral collisions on the $1/I_{\rm p}$ scaling of heat-load width is shown not to be significant.  Although inclusion of  poloidally uniform anomalous transport results in a deviation from the $1/I_{\rm p}$ scaling, the inclusion of the anomalous transport that is driven by ballooning-type instabilities results in recovering the neoclassical  $1/I_{\rm p}$  scaling. The Bohm or Gyro-Bohm scalings of anomalous transport does not strongly affect the dependence of the heat-load width on plasma current. The inclusion of anomalous transport, in general, results in widening the width of neoclassical divertor heat-load and enhances the neoclassical heat-load fluxes on the divertor plates. Understanding heat transport in the tokamak scrape-off layer plasmas is important for strengthening the basis for predicting divertor conditions in ITER.\footnote{Submitted to Physics of Plasmas}

\end{abstract}
%in the scrape-off layer (SOL) region
%The neoclassical divertor heat-load width is found to be broader for smaller
% plasma currents $I_{\rm p}^{-0.8}$ in DIII-D discharges.
\maketitle
\section{Introduction}
Insight regarding heat transport in tokamak scrape-off layer plasmas is important for strengthening the basis for envisaging divertor conditions in existing and future tokamak devices~\cite{herrmann02,labombard11,brunner13,eich13,maingi14,polevoi15,bourdelle15}. The XGC0 particle in cell 4D gyro-kinetic torodially symmetric code with specified background diffusion coefficients can be used to study orbit width effects on the scrape-off layer~\cite{chang2004}. In the research presented in this paper, the XGC0 code is used to investigate kinetic neoclassical behavior of the heat fluxes on the divertor plates in Alcator \hbox{C-Mod} and DIII-D tokamaks employing realistic divertor geometry. It is important for understanding the dependencies that affect the divertor heat-load fluxes in planning experiments and in developing new models to describe the physics associated with the scrape-off layer region.

The neoclassical divertor heat-load fluxes are computed and changes in the fluxes associated with the introduction of the effects of neutral collisions and anomalous transport are examined.  In particular, the dependence of the divertor heat-load width on the plasma current is studied. The anomalous transport in the XGC0 neoclassical code is applied in three different ways: i) Within 45 degrees from the midplane, ii) uniformly for all poloidal angles and iii) using the Bohm and gyro-Bohm type scalings.

The discharges examined include an Alcator C-Mod discharge and a series of four DIII-D discharges that represent a plasma current scan~\cite{groebner2009}. The Alcator C-Mod discharge 1100212024 was a discharge in an Alcator \hbox{C-Mod/DIII-D} similarity campaign. The total plasma current in the Alcator C-Mod discharge is 0.9~MA and the toroidal magnetic field is 5.4~T. The discharge is an enhanced D$_\alpha$ (EDA) high-confinement discharge with auxiliary heating of approximately 3.9~MW provided by ICRH. In the DIII-D discharges, the total plasma current is varied from 0.51 MA to 1.50 MA with a fixed toroidal magnetic field ($B_{\rm T} \approx 2.1 {\rm T}$), plasma triangularity ($\delta \approx 0.55$), and normalized toroidal beta ($\beta_{\rm n}\approx 2.1- 2.4$). The total plasma current and auxiliary heating power are given in the Table~\ref{table}. The plasma density at the top of the pedestal varies from approximately $4.5 \times 10^{19} \textrm{m}^{-3}$ in the high plasma density DIII-D discharge, 132016, to $2.5 \times 10^{19} {\rm m}^{-3}$ in the low plasma density DIII-D discharge, 132018.

\begin{table*}
\begin{center}
\begin{tabular}{|c|c|c|c|}
\hline
DIII-D Discharges & EFIT time            & Plasma current             &
Auxillary  heating  power \\
    \             & msec                  &  MA                 &  MW\\
\hline \hline
132016  & 3023 & 1.50 & 8.12\\
\hline
132014 & 3023 & 1.17 & 7.36\\
  \hline
132017 & 2998 & 0.85 & 8.50\\
  \hline
132018 & 1948 & 0.51 & 7.10\\
 \hline
\end{tabular}
\end{center}\vskip-0.15in
\caption{\it \baselineskip 8pt The DIII-D discharges analyzed in this study and
their parameters.}\label{table}
\end{table*}

The divertor heat-load width is found to be broader in the simulations of discharges with lower plasma current. The neoclassical effects are essential to explain the $1/I_{\rm p}$ scaling that has been observed in tokamak discharges. However, the $1/I_{\rm p}$ scaling in Alcator C-Mod discharge is found to be weaker as compared to the scaling observed in the DIII-D discharges. This result is consistent with experimental observations~\cite{hill1992,lasnier1998}. The effects of neutral collisions on the neoclassical heat-load width are compared to the neoclassical heat-load width computed without the inclusion of neutral collisions. Neutral collisions at the edge of the plasma are shown to have a limited effect on the divertor heat-load width. It is found that the poloidally uniform distribution of anomalous flux can significantly alter the $1/I_{\rm p}$ scaling. The gyro-Bohm scaling of anomalous transport has somewhat weaker effect on the divertor heat-load width $1/I_{\rm p}$ scaling than does the Bohm scaling of anomalous transport.

The manuscript is organized as follows: In Sec.~\ref{XGC0}, following a brief description of XGC0 neoclassical code, the computation of XGC0 heat fluxes on the divertor plates is presented.  Sec.~\ref{neo} contains a description of the neoclassical prediction for the Alcator C-Mod and DIII-D divertor heat fluxes. The effects of neutral collisions on the divertor heat-load width prediction are presented in Sec.~\ref{neut}. The anomalous transport model in the XGC0 code is discussed in Sec.~\ref{prof}. The effects of anomalous transport on the width of heat flux on the divertor are presented in Sec.~\ref{anoma} and the results are summarized in Sec.~\ref{Summary}.

\section{Computation of XGC0 Neoclassical heat fluxes~\label{XGC0}}

%---------Figure 1-------------
\begin{figure}
\centering{}%
\includegraphics[height=2.50in,width=3.2in,clip]{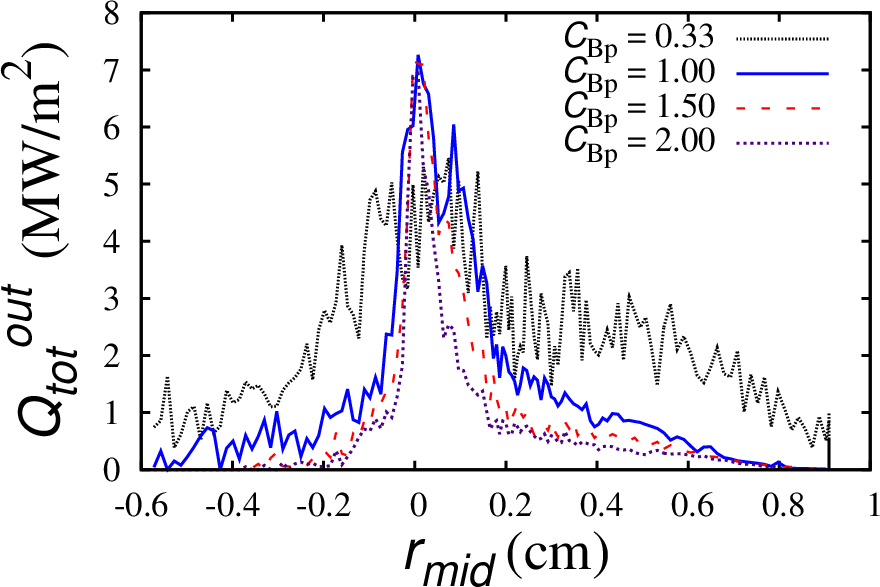}
\caption{The XGC0 neoclassical heat flux to the outer divertor plate ($Q_{\rm tot}^{\rm out}$) versus distance mapped to midplane ($r_{\rm mid}$) for the Alcator C-Mod discharge 1100212024 computed using plasma current scaling factors $C_{\rm Bp} = 0.33,\ 1.00,\ 1.50,\ {\rm and}\ 2.00$.
\label{figfluxcmod}}
\end{figure}

%---------Figure 2-------------
\begin{figure*}
\centering{}%
\begin{tabular}{llll}
\includegraphics[height=2.50in, width=3.20in,clip]{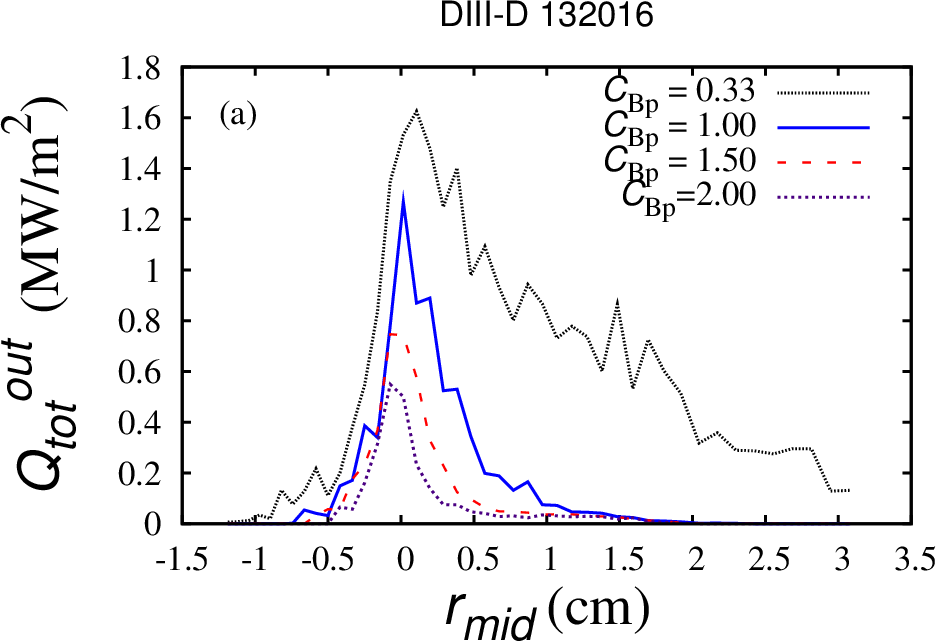} &
\includegraphics[height=2.50in,width=3.20in,clip]{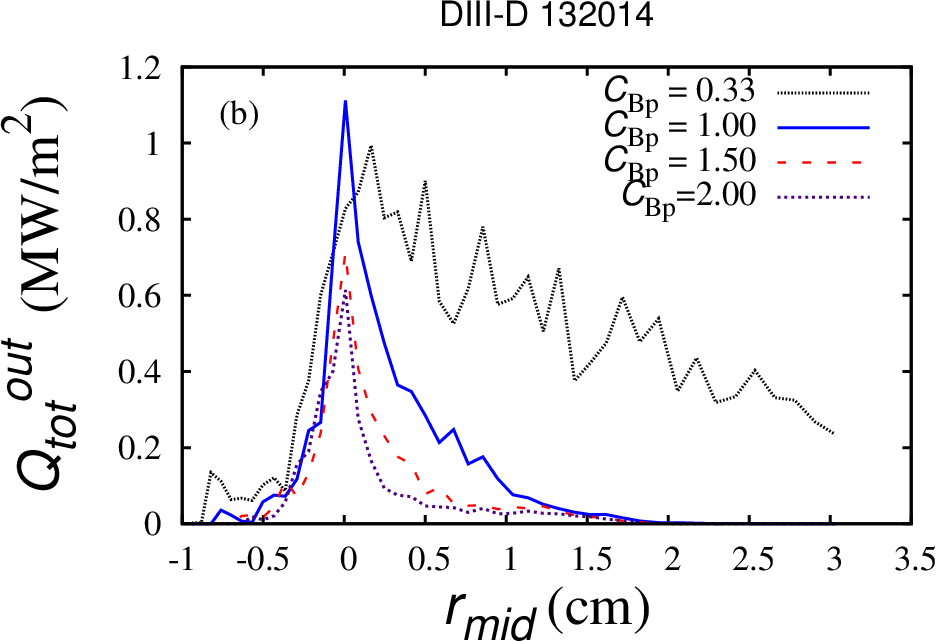} \\
\\
\includegraphics[height=2.50in,width=3.20in,clip]{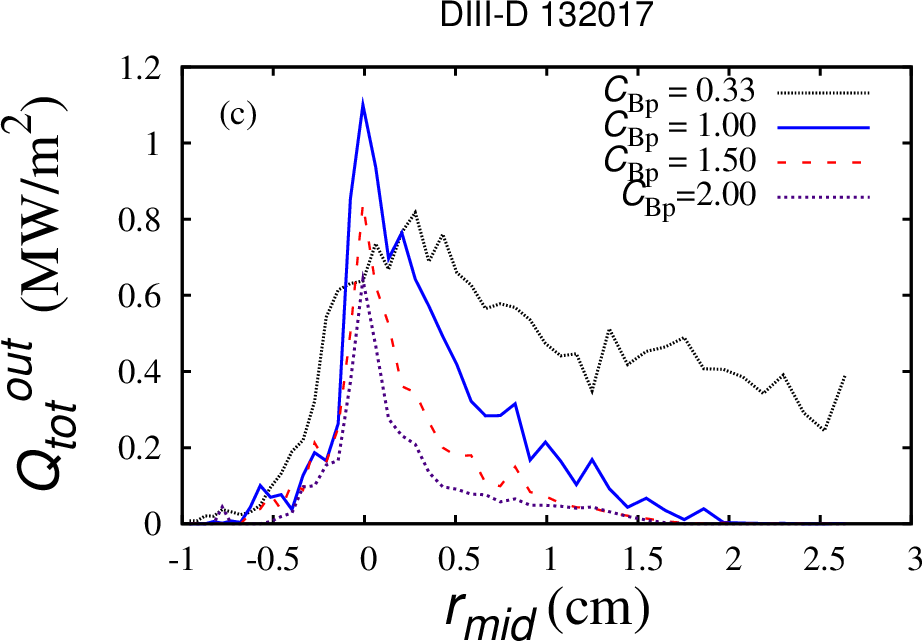} &
\includegraphics[height=2.50in,width=3.20in,clip]{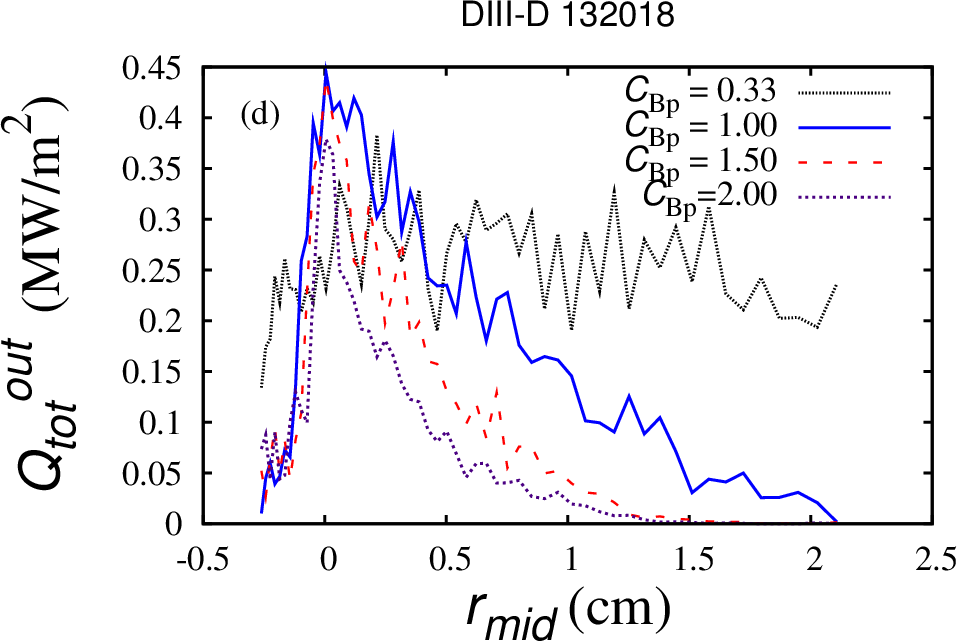} \\
\end{tabular}
\caption{  \label{fig:fluxd3d} The XGC0 neoclassical heat fluxes to the outer divertor plate versus distance mapped to midplane ($r_{\rm mid}$) for four DIII-D discharges.  The plasma current scaling factors used are $C_{\rm Bp} = 0.33$,  $C_{\rm Bp} = 1.0$, $C_{\rm Bp} = 1.50$ and $C_{\rm Bp} = 2.0$.  The results shown in panel (a) are for discharge 132016 with $I_{\rm p} = 1.50\,{\rm MA}$; in panel (b) for discharge 132014 with $I_{\rm p} = 1.17\,{\rm MA}$; in panel (c) for discharge 132017 with $I_{\rm p} = 0.85\,{\rm MA}$; and in panel (d) for discharge 132018 with $I_{\rm p} = 0.51\,{\rm MA}$.}
\end{figure*}

In the XGC0 neoclassical code~\cite{chang2004,ku04}, the gyrokinetic plasma ion and electron guiding centers evolve in time within a five-dimensional phase space using a realistic magnetic equilibrium and limiter geometry. The code uses cylindrical coordinates so that the separatrix and X-point region can be easily included in the simulation domain. In the XGC0 code, the motion of charged and neutral particle are followed, and the effect of collisions are taken into account using a Monte Carlo approach. The XGC0 code includes models for the source of neutrals at the wall and solves the Hamiltonian guiding center equations of motion described in Ref.~\citenum{white90,boozer84,littlejohn85}.

A combination of cylindrical and field-line-following magnetic coordinates are used in the XGC0 code. Magnetic coordinates have the advantage of accurately treating physical phenomena that are aligned with the magnetic field lines. Cylindrical coordinates have the advantage of being able to treat the open magnetic surfaces of the scrape-off-layer and the closed magnetic surfaces of the plasma core with equal accuracy and without the divergence associated with magnetic coordinates at the separatrix.

The XGC0 neoclassical heat fluxes to the divertor plates ($Q_{\rm tot}^{\rm out}$) are computed for the Alcator C-Mod discharge and for the four DIII-D discharges.  In order to test the sensitivity of the heat flux with regard to plasma current, a multiplicative scaling factor $C_{\rm Bp}$ is introduced. The scaling factor $C_{\rm Bp}$ is an internal numerical multiplier introduced in the XGC0 code in order to scale the equilibrium poloidal flux.  The toroidal flux is not modified. Thus, the amplification factor $C_{\rm Bp}$  can be also considered as a scaling factor for the total plasma current.

In Fig~\ref{figfluxcmod}, the computed neoclassical heat flux to the outer divertor plate $Q_{\rm tot}^{\rm out}$ is shown for the Alcator C-Mod discharge 1100212024, a discharge that was a part of Alcator C-Mod/DIII-D similarity campaign. The plasma current scaling factors, $C_{\rm Bp}$, ranged from 0.33 to 2.0. The value of $Q_{\rm tot}^{\rm out}$ is found to be large in the region close to the separatrix. A lower peak value of $Q_{\rm tot}^{\rm out}$ and a broader width is obtained when $C_{\rm Bp} = 0.33$ than obtained with higher values of $C_{\rm Bp}$.  It is seen that as the $C_{\rm Bp}$ scaling factor is increased from 1.0 to 2.0, the width of the heat flux to the divertor plate continues to decrease although the change is small relative to the change observed in the results obtained for the D3D discharges indicating that the Alcator C-Mod discharge has a weaker scaling of the divertor heat-load width with plasma current.

In Fig.~\ref{fig:fluxd3d}, results for the heat flux, $Q_{\rm tot}^{\rm out}$, versus $r_{\rm mid}$ are presented for the four DIII-D discharges considered. The four discharges differ primarily in terms of the total plasma current, as shown in Table~\ref{table}. It is observed that an increase in the $C_{\rm Bp}$ scaling factor results in a decrease in the width of the heat-load on the divertor plate. For values of $C_{\rm Bp} \geq 1.0$  the peak value neoclassical heat flux to the divertor decreases as $C_{\rm Bp}$ increases.  This is more pronounced in the simulation results for the higher current DIII-D discharge, 132017, shown in Fig.~2(a).  The magnitude of the neoclassical heat flux is found to be generally smaller for the low plasma current discharges. Similar to the result for the Alcator C-Mod discharge, the heat flux in the DIII-D discharges is also found to be large in the region close to the separatrix. However, the magnitude of the neoclassical heat flux close to the separatrix in the DIII-D discharges is smaller than the corresponding magnitude obtained in the simulation Alcator C-Mod discharge (shown in Fig~\ref{figfluxcmod}).
\subsection{Sensitivity of neoclassical prediction of divertor heat-load width on plasma current \label{neo}}
%---------Figure 3-------------
\begin{figure*}
\centering{}%
\begin{tabular}{ll}
\includegraphics[width=3.20in,clip]{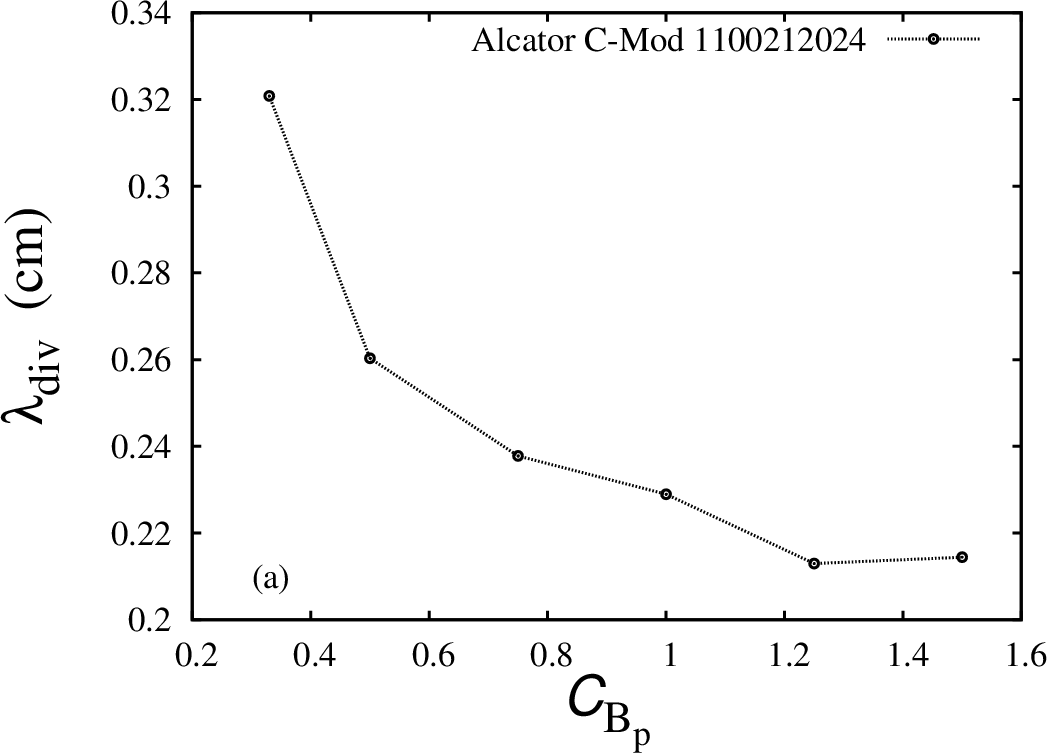} &
\includegraphics[width=3.20in,clip]{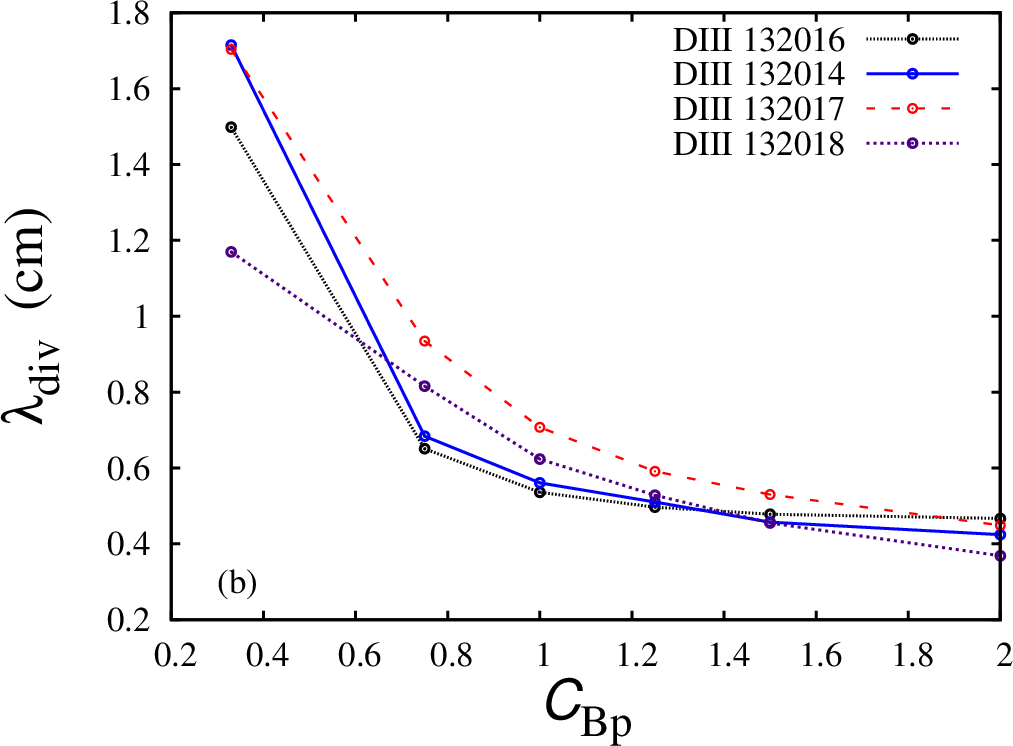} \\
\end{tabular}
\caption{\label{lambda-neo} The XGC0 neoclassical heat-load width ($\lambda_{\rm div}$) versus a plasma current scaling factor $C_{\rm Bp}$ for a) the Alcator C-Mod discharge and b) for the four DIII-D discharges.}
\end{figure*}
In Fig.~\ref{lambda-neo}, the computed XGC0 neoclassical heat-load width ($\lambda_{\rm div}$) is plotted versus the plasma current scaling factor ($C_{\rm Bp}$) for the Alcator C-Mod discharge and for the four DIII-D discharges considered. It is found that the heat-load width decreases with increasing plasma current, that is $\lambda_{\rm div}\propto I_{\rm p}^{\rm -\alpha}$. In the Alcator C-Mod discharge, the dependence of $\lambda_{\rm div}$ on $I_{\rm p}$ is weaker as compared to the DIII-D discharges, shown in Fig.~\ref{lambda-neo}b. For the base case ($C_{\rm Bp} = 1.0$), the neoclassical divertor heat-load width in the Alcator C-Mod discharge is found to be approximately 2.3 mm, which is about 20\% below the experimentally observed value~\cite{labombard11} of approximately 3 mm. As shown below in Sec.~\ref{anoma}, the anomalous effects typically increase the divertor heat-load width bringing the simulation and experimental results closer to one-another.

In Fig.~\ref{lambda-neo}b it is seen that the change in heat-load width with increasing scaling factor, $C_{\rm Bp}$, is not the same for the four DIII-D discharges.  For $C_{\rm Bp}\ge 0.75$, the decrease in diverter heat-load width with increasing $C_{\rm Bp}$ for the two DIII-D discharges with higher plasma density, 132014 and 132016, is less than the decrease for the two DIII-D discharges with lower plasma density, 132017 and 132018.  As indicated below, this difference may, in part, be due to the omission of the effects of collisions.
\subsection{Effect of neutral collisions on divertor heat-load width}
\label{neut}
\begin{figure}
\centering{}%
\includegraphics[width=3.20in,clip]{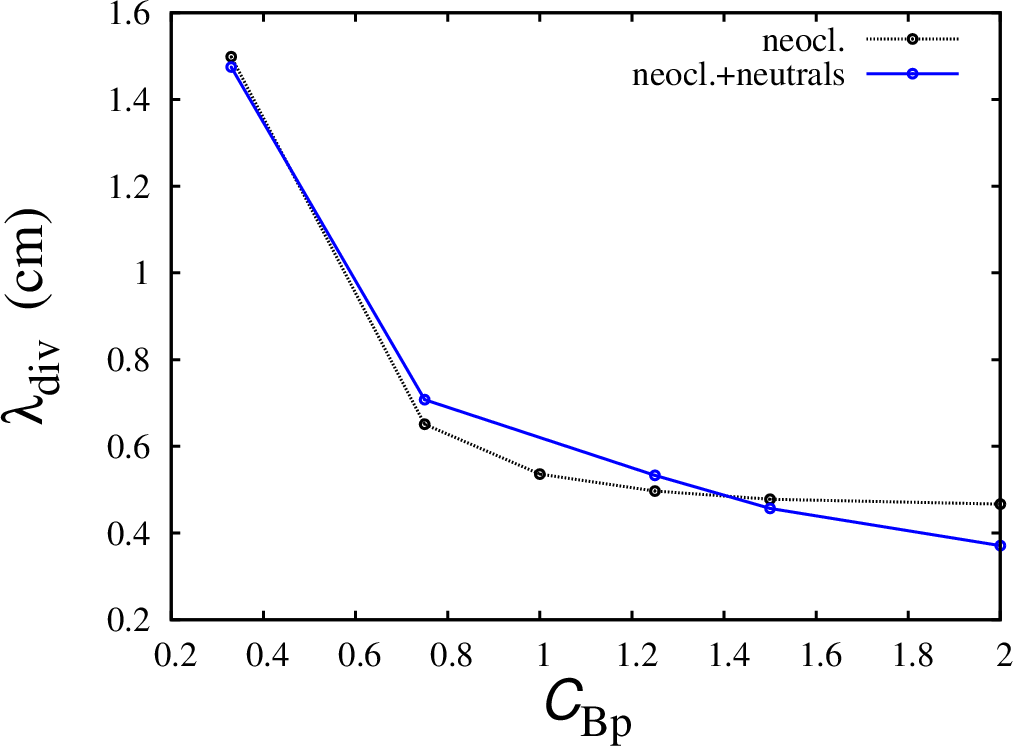}
\caption{\label{fig:lambda-neut} The XGC0 neoclassical heat-load width for  with and without the neutral collisions as functions of plasma current scaling factor $C_{\rm Bp}$ for DIII-D discharge 132016.}
\end{figure}
In Fig.~\ref{fig:lambda-neut}, the XGC0 neoclassical heat-load width in the presence of neutrals is shown as a function of plasma current scaling factor $C_{\rm Bp}$ for DIII-D discharge 132016. The dotted curve shows the dependence of the width of neoclassical heat-load when the effects of neutral collisions are not include, the solid curve shows the heat load width, when the neutral collisionsare included. The dependence of the divertor heat-load width is weakly affected by neutral collisions especially at lower plasma currents (lower values of $C_{\rm
Bp}$) where neoclassical effects are found to be strong.

\section{Finding anomalous transport profiles which yield XGC0 predicted plasma
profiles consistent with experimental profiles \label{prof}}
Anomalous transport in the plasma edge region has been analyzed previously, and it has been found that particle pinches may have an important role in the pedestal region~\cite{rensink1993, stacey2010}. Experimental observations and some analytical models suggest that anomalous transport in scrape-off layer may be intermittent and significantly larger than in the pedestal region. Studies of the anomalous transport in the plasma edge region are typically based on the analysis of experimental data and include a robust model for neoclassical transport. Since particle and thermal fluxes obtained from analysis of experimental data include contributions from both the anomalous and neoclassical transport, it is necessary to deduct the neoclassical transport in order to obtain the correct anomalous fluxes and effective anomalous diffusivities.

Although theory-based models for anomalous transport are available in the XGC0 code, a goal of the study described in this section is to derive the anomalous effective diffusivities that can reproduce the experimental profiles when these diffusivities are utilized in the neoclassical kinetic XGC0 code. It is found that strong pinches in all channels of anomalous transport are necessary to reproduce the experimental profiles. The diffusivities that reproduce the experimental profiles are then used in the computation of the divertor heat-load fluxes.
%---------Figure 5-------------
\begin{figure}
	\includegraphics[width=3.20in,clip]{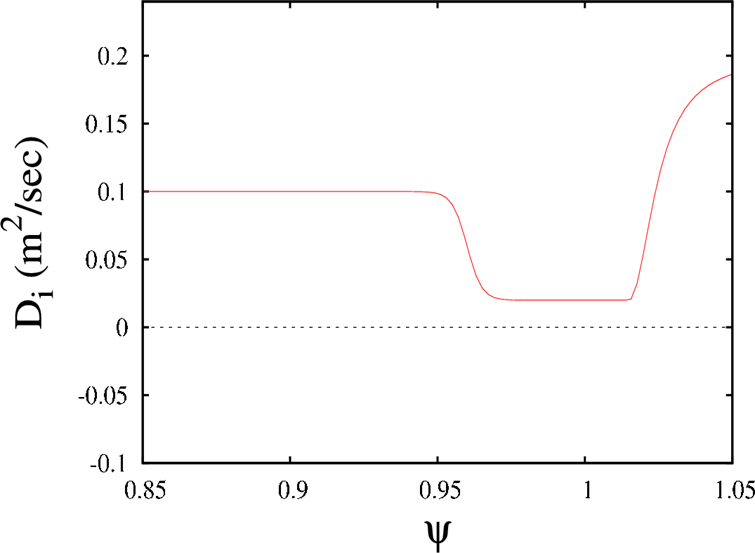}
%\vspace{-0.2in}
\caption{\label{di-profile}  A typical effective particle diffusivity versus normalized poloidal flux ($\psi$) used in the simulations of Alcator C-Mod and DIII-D discharges. The particle diffusivity profile is selected to reproduce the experimental particle density profile, the solid line curve in Fig.~\ref{n-profile}.  }
\end{figure}
%To begin with, anomalous diffusivity profiles are kept fixed and assumed to be
% poloidally uniform for each discharge in all the plasma current scans.
%---------Figure 6-------------
\begin{figure}
\includegraphics[width=3.2in,clip]{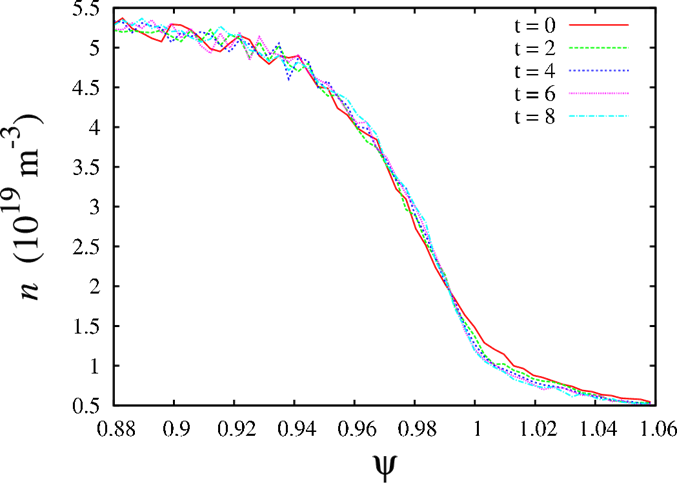}
\caption{\label{n-profile}  The XGC0 simulation results of the plasma
density as a function of normalized poloidal flux ($\psi$) in the DIII-D
discharge 132016 for the first eight ion transit periods. The solid line indicates the experimental
plasma density profile.}
\end{figure}

The XGC0 code utilizes an enhanced random walk model for the anomalous thermal and particle transport. Unlike the standard random walk model that uses the same diffusion coefficients for all transport channels, the XGC0 code distinguishes between the particle and thermal transport. The model is based on the Monte-Carlo
method for the advection-diffusion equation in the phase space for the anomalous transport:
\begin{equation}
\frac{\partial f}{\partial t}=-\nabla \cdot {\bf\Gamma}_{\rm A},
\end{equation}
where the anomalous flux, ${\bf\Gamma}_{\rm A}$, is defined as
\begin{equation}
{\bf\Gamma}_{\rm A}=-D_{\rm A}\nabla f+\left[\left(D_{\rm A}-\frac{2}{3}\chi_{\rm A}\right)
\left(\frac{E}{T}-\frac{3}{2}\right)\frac{\nabla T}{T}\right]f
\end{equation}
The coefficients $D_{\rm A}$ and $\chi_{\rm A}$ are the anomalous particle and thermal transport coefficients respectively. These coefficients can be obtained from theory-based models or prescribed as an input profiles in the XGC0 code. In this model, the Monte-Carlo particles advance in $RZ$ space depends on both anomalous particle and thermal transport coefficients:
\begin{widetext}
\begin{eqnarray}
R &\rightarrow& R + \left[ \frac{1}{R}\frac{\partial}{\partial R} \left(D_{\rm A}R\right) +
\left[\left(D_{\rm A}-\frac{2}{3}\chi_{\rm A}\right)
\left(\frac{E}{T}-\frac{3}{2}\right)\frac{\nabla T}{T}\right]\cdot \hat{e}_{Z}\right]\Delta t\pm \sqrt{2D_{\rm A}\Delta t}\\
Z &\rightarrow& Z + \left[ \frac{\partial D_{\rm A}}{\partial Z}  +
\left[\left(D_{\rm A}-\frac{2}{3}\chi_{\rm A}\right)
\left(\frac{E}{T}-\frac{3}{2}\right)\frac{\nabla T}{T}\right]\cdot \hat{e}_{Z}\right]\Delta t\pm \sqrt{2D_{\rm A}\Delta t}
\end{eqnarray}
\end{widetext}

%---------Figure 7-------------
\begin{figure*}[t]
\centering{}%
\begin{tabular}{ll}
\includegraphics[height=2.50in,width=3.20in,clip]{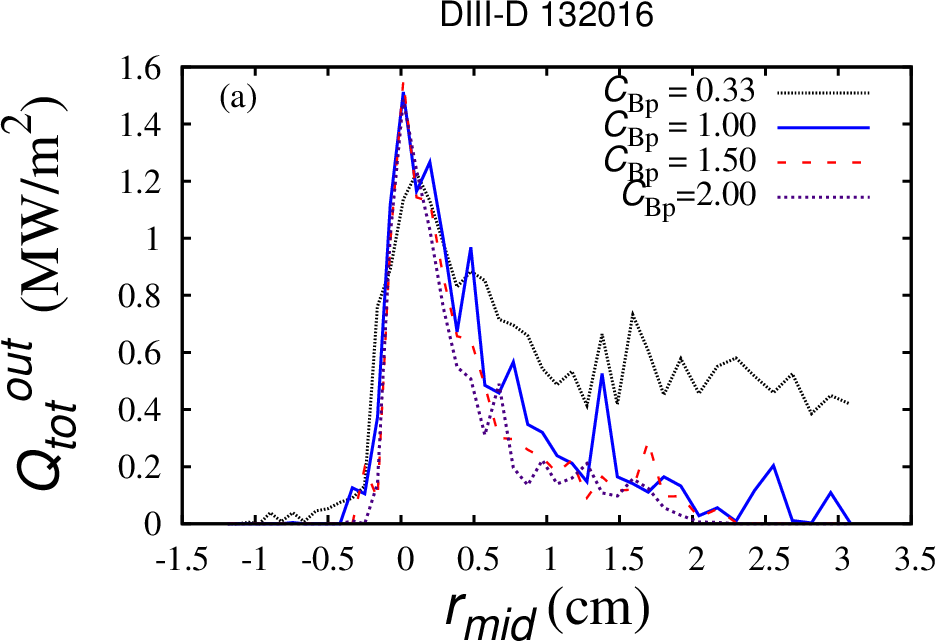} &
\includegraphics[height=2.50in,width=3.20in,clip]{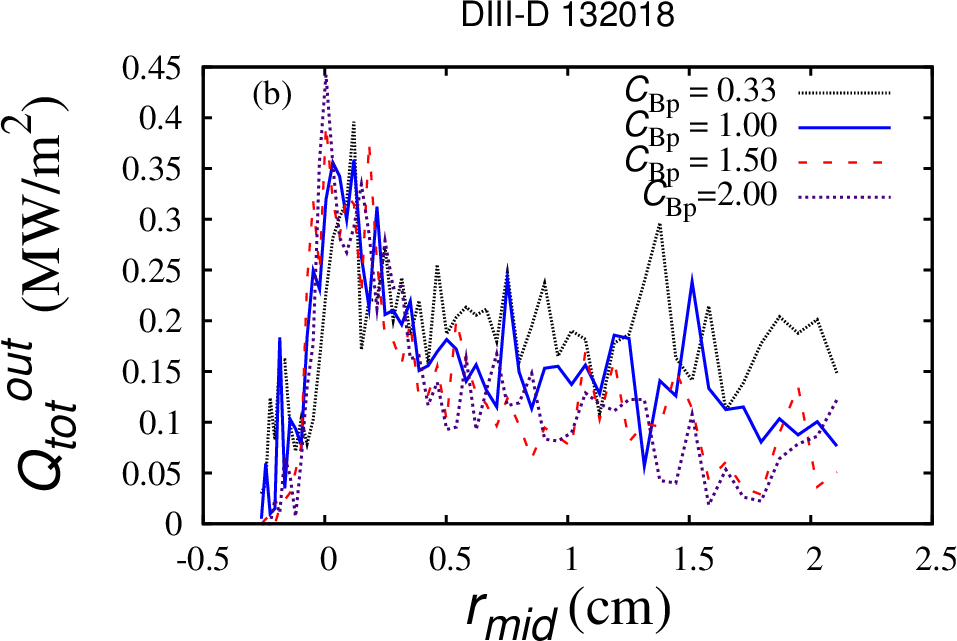} \\
\end{tabular}
\caption{\label{fig:fluxd3da} DIII-D heat fluxes profiles, with the anomalous transport localized near the midplane, versus distance mapped to midplane ($r_{\rm mid}$) for four values of plasma current scaling factor $C_{\rm Bp}$. Panel (a) contains the results for the high plasma current ($I_{\rm p} = 1.50\, {\rm MA}$) DIII-D discharges 132016. Panel (b), the low plasma current ($I_{\rm p} = 0.51\, {\rm MA}$)  DIII-D discharge 132018. }
\end{figure*}

In XGCO code, three regions of constant diffusivites are implemented.  The regions of constant diffusivities  are separated by two narrow transitional regions that use tanh-fit. The levels of anomalous transport in the three regions, the locations of transitional regions and their widths are adjustable parameters. The particle diffusivity profiles as well as the electron and ion thermal diffusivity profiles are independent of one another. These profiles are adjusted to find a steady state solution that reproduces the experimental profiles. The anomalous diffusivity profiles are selected so that the resulting density and temperature profiles remains close to the experimental profiles for at least eight ion transit periods. For example, the effective diffusivity shown in Fig.~\ref{di-profile} is such that when it is combined with the neoclassical transport computed with XGC0, the experimental profiles, shown in Fig.~\ref{n-profile}, are reproduced. It has been found that particle and thermal pinches play an important part in the pedestal regions of all discharges analyzed in this study.

\section{Effect of anomalous transport on the width of divertor heat fluxes
\label{anoma}}
The anomalous transport in XGC0 neoclassical code applied in three different ways: i) Within 45 degrees from the midplane, ii) uniformly for all poloidal angles iii) Bohm and gyro-Bohm type scaled transport.

The heat flux profiles, with anomalous transport within 45 degrees from the midplane, are shown in Fig.~\ref{fig:fluxd3da} for DIII-D discharges 132016 and 132018 for four values of the plasma current scaling factor. In these simulations, the implementation of the anomalous transport in the region within $45^{\circ}$ from the midplane implies that the modes which contribute to the anomalous transport at the plasma edge are ballooning in nature. The heat fluxes profiles shown in Fig.~\ref{fig:fluxd3da} are found to be somewhat broader in comparison with the heat fluxes obtained with only neoclassical transport (see Panel a and Panel d of Fig.\ref{fig:fluxd3d}). Thus, the introduction of anomalous transport results in widening the divertor heat-load width. The ballooned anomalous loss (outboard midplane), even without an $I_{\rm p}$ dependence, recovers neoclassical plasma current scaling proportional to $I_{\rm p}^{-0.6}$. This scaling in the presence of anomalous transport implies that the effects associated with anomalous transport that is localized around the outboard midplane is not significant.

%---------Figure 8-------------
\begin{figure}
\centering{}%
\includegraphics[width=3.20in,clip]{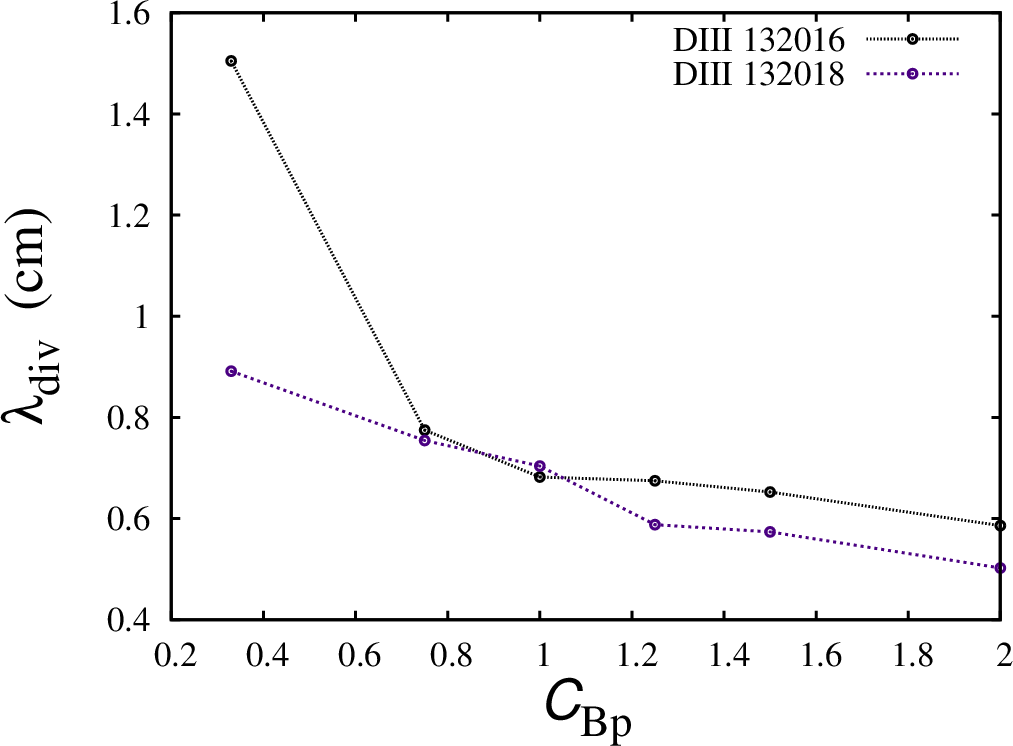}
\caption{\label{fig:lambda-anom1}The divertor heat-load width as a function of the plasma current scaling factor $C_{\rm Bp}$ for DIII-D discharges 132016 ($I_{\rm p} = 1.50\, {\rm MA}$) and 132018 ($I_{\rm p} = 0.51\, {\rm MA}$).  The simulations are carried out using the assumption that the anomalous transport is localized near the midplane. Poloidally uniform and $I_{\rm p}-$independent anomalous transport can mask the neoclassical scaling. This can be seen by comparing these results with the neoclassical divertor heat-load width for the discharges 132016 and 132018 plotted in Fig~\ref{lambda-neo}.}
\end{figure}

In Fig.~\ref{fig:lambda-anom1}, the divertor heat-load width is plotted as a function of plasma current scaling factor $C_{\rm Bp}$ for DIII-D discharges 132016 and 132018.  In carrying out these simulations the assumption is made that the anomalous transport is localized near the midplane. The divertor heat-load width decreases with increasing $C_{\rm Bp}$ even in the presence of ballooned anomalous transport. This implies that the neoclassical dependence of the divertor heat-load width on the total plasma current is preserved.

%---------Figure 9-------------
\begin{figure}
\centering{}%
\includegraphics[height=2.50in,width=3.20in,clip]{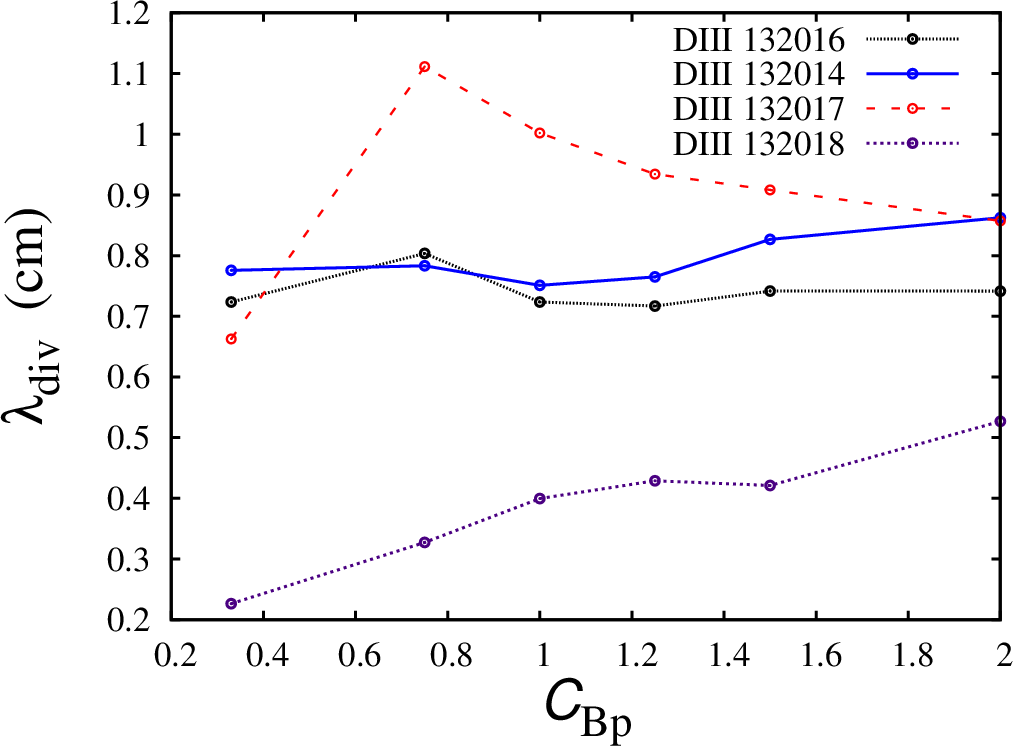}
\caption{\label{fig:lambda-anom2} The heat-load width ($\lambda_{\rm div}$) versus a plasma current scaling factor $C_{\rm Bp}$ for four DIII-D discharges. The anomalous transport applied uniformly for all poloidal angles. A poloidally uniform, $I_{\rm p}$ independent, anomalous transport eliminates the neoclassical
$I_{\rm p}$ scaling behavior.}
\end{figure}

The divertor heat-load width is plotted in Fig.~\ref{fig:lambda-anom2} as a function of the current scaling factor for the four DIII-D discharges when the simulations are carried out with the anomalous transport applied uniformly for all poloidal angles. It is found that for all four DIII-D discharges the simulated heat-load width no longer decreases with increasing plasma current.  A poloidally uniform, $I_{\rm p}$ independent, anomalous transport eliminates the neoclassical $I_{\rm p}$ scaling behavior.

%---------Figure 10-------------
\begin{figure}
\centering{}%
\includegraphics[height=2.50in,width=3.20in,clip]{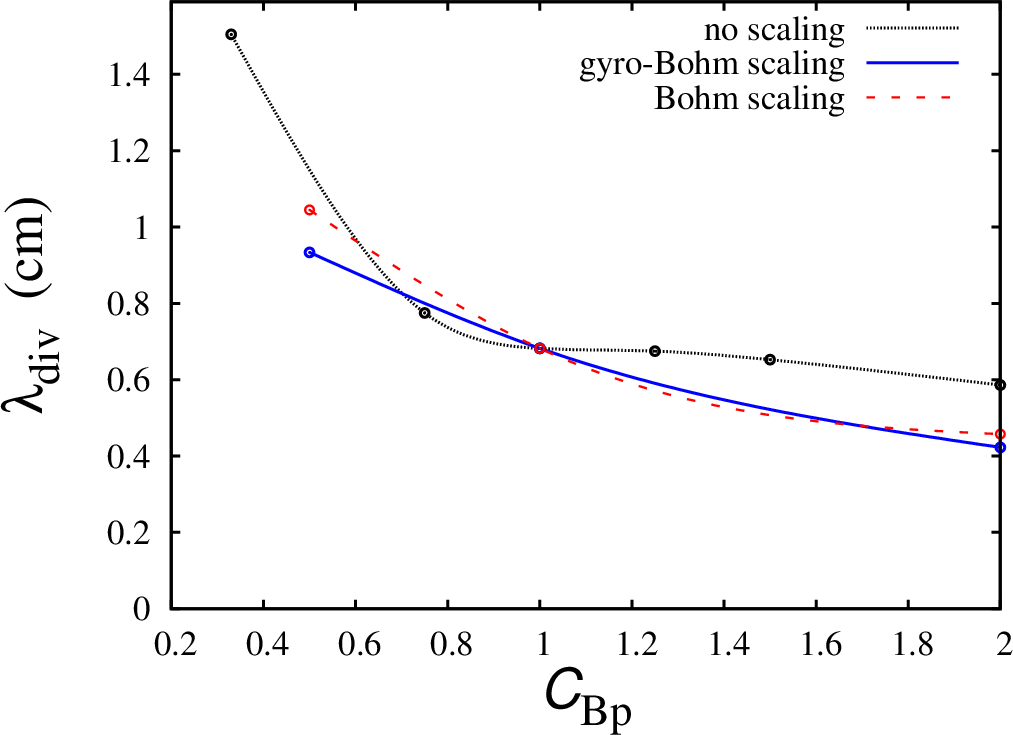}
\caption{\label{fig:fluxcmod}  The effect of Bohm and gyro-Bohm scalings of anomalous transport on the predictions of the heat-load width for the DIII-D discharge 132016. The gyro-Bohm and Bohm scalings for anomalous transport have been implemented in XGC0 neoclassical code.  The magnetic field dependent  Bohm and gyro-Bohm scalings of anomalous transport also recovers the neoclassical type $I_{\rm p}^{- \alpha}$ scaling.}
\end{figure}

The effects of Bohm and gyro-Bohm scalings of anomalous transport on the predictions of the divertor heat-load width for the high current and high density DIII-D discharge 132016 are shown in Fig.~\ref{fig:fluxcmod}. The heat-load width computed using Bohm and gyro-Bohm scalings of anomalous transport is compared with the heat-load width when these scalings are not used. It is found that the magnetic field dependent Bohm and gyro-Bohm type scalings of anomalous transport recover the neoclassical type $I_{\rm p}^{- \alpha}$ scaling. However, both Bohm and gyro-Bohm scalings result in narrower heat-load widths at large and small values of plasma current as compared to the heat-load width when these scalings are not applied in the simulations.

\section{Summary \label{Summary}}
A goal of this study is to improve the understanding of scrape-off layer thermal transport. Understanding the physical effects that contribute to divertor heat-load fluxes is important in planning experiments, designing future tokamaks, and developing new models for the scrape-off layer region. The heat fluxes and heat-load width in scrape-off layer are computed using the kinetic neoclassical XGC0 code. The guiding-center neoclassical transport code XGC0 includes realistic X-point geometry, a neutral source model, collisions, and diffusion models for anomalous transport. 

An Alcator C-Mod discharge, which was a part of Alcator C-Mod/DIII-D similarity campaign, and a series of four DIII-D discharges that represent plasma current scan are considered. The neoclassical divertor heat-load fluxes and a relation for the dependence of the divertor heat-load width on the plasma current are obtained in the simulations carried out. In examining the dependence of divertor heat-load width on plasma current, effects of neutral collisions and anomalous transport are taken into account. Changes in the neoclassical divertor heat-load fluxes associated with the introduction of the neutral collisions and anomalous transport are described. Anomalous transport in XGC0 neoclassical code is applied in three different ways: i) Within 45 degrees from the midplane, ii) uniformly over all poloidal angles and iii) with Bohm and gyro-Bohm type scalings. The effective particle and thermal diffusivities that are used in the simulations are such that when they are combined with the neoclassical transport computed with XGC0 code, the experimental profiles are reproduced. The profiles for the particle diffusivity as well as for the electron and ion thermal diffusivities are independent of one another and are adjusted to find a steady state solution so that the resulting profiles remains close to the experimental profiles for at least eight ion transit periods.

The simulations of Alcator C-Mod and DIII-D discharges indicate that the variation of neoclassical heat-load width on the outer divertor plates with plasma current is given by $I_{\rm p}^{- \alpha}$ where $\alpha$ is approximately 0.3 for the Alcator C-Mod discharge and 0.8 for DIII-D discharges.  The computed heat-load width in the simulation of Alcator C-Mod discharge is roughly consistent with experimental measurement of approximaltely 3~mm for this discharge~\cite{labombard10}. The computed value of 2.3~mm is obtained in a neoclassical limit when the anomalous effects are excluded. Inclusion the amomalous effects results in a wider width and brings the head-load width prediction closer to the experimental value. The weaker dependence on $I_{\rm P}$ found for Alcator C-Mod can be attributed to higher collisionality in Alcator C-Mod relative to DIII-D. In general, the heat flux is found to be large in the area close to the separatrix and the heat flux profiles become broader for lower values of plasma current. The neutral collisions do not significantly modify the dependence of the neoclassical heat-load width on plasma current.  The dependence of the neoclassical heat-load width on plasma current is modified when the anomalous transport is applied uniformly for all poloidal angles.  However, The neoclassical dependence of the divertor heat-load width on the plasma current is preserved when the anomalous transport is applied in the region within $45^{\circ}$  from the midplane. The Bohm and gyro-Bohm scalings of anomalous transport in XGC0 simulations results in narrower divertor heat-load widths that exhibit the $I_{\rm p}^{- \alpha}$ dependence.
\section{Acknowledgements}
This material is based upon work supported by the U.S. Department of Energy, Office of Science, under Award Numbers DE-SC0006629, DE-SC0008605, DE-SC0012174, and DE-FG02-92-ER54141.
\bibliography{papref}
\end{document}